\documentclass[12pt,twoside]{article}
\usepackage{amsmath,amsfonts,amssymb,amsthm}
\usepackage{bbm}
\usepackage{graphicx}
\usepackage[T1]{fontenc}

\usepackage{color}
\newcommand{\grad}{\nabla}

\newcommand{\laplace}{\Delta}

\renewcommand{\div}{\grad\cdot}

\newcommand{\R}{\mathbf{R}}

\def\Nu{{\mathit{Nu}}}
\def\Ra{{\mathit{Ra}}}
\def\Re{{\mathit{Re}}}
\def\Pr{{\mathit{Pr}}}
\def\Gr{{\mathit{Gr}}}

\newcommand{\const}{\mbox{\em const}}

\newcommand{\eps}{\varepsilon}
\newcommand{\la}{\langle}

\newcommand{\ra}{\rangle}

\begin{document}
%\changefont{ppl}{m}{n}

\title{Scaling bounds on dissipation in turbulent flows}
\author{Christian Seis\footnote{Institut f\"ur Angewandte Mathematik, Universit\"at Bonn, Endenicher Allee 60, 53115 Bonn, Germany}}

\maketitle

\begin{abstract}
We present a new rigorous method for estimating statistical quantities in fluid dynamics such as the (average) energy dissipation rate directly from the equations of motion. The method is tested on shear flow, channel flow, Rayleigh--B\'enard convection and porous medium convection.
\end{abstract}

\section{Introduction}

One of the most fascinating features in turbulent flows is the emergence of complicated chaotic structures involving a wide range of length scales behind which ``typical'' flow patterns are still recognizable.
The state of motion is too complex to allow for a detailed description of the fluid velocity and experimental or numerical measurements of certain system quantities appear disorganized and unpredictable. Yet, some statistical properties are reproducible \cite{Frisch95}. One of the challenges in theoretical fluid dynamics is thus the derivation of quantitative statements on turbulent flows. Many of the approaches consist of various approximation procedures, the imposition of physically motivated but ad hoc assumptions (e.g., ``closure'') or the introduction of scaling hypotheses. Rigorous results beginning with the equations of motion directly are therefore indispensable for checking the validity of the imposed simplifications and for justifying secondary models and theories.

The idea of extracting information about driven turbulent flows via bounds on physical quantities through mathematically justifiable operations and without imposing ad hoc assumptions goes back to the pioneering works of Malkus, Howard and Busse in the 1950s and 60s \cite{Malkus54,Howard63,Howard72,Busse70,Busse79}. These authors applied variational approaches for the derivation of bounds on the energy dissipation rate in models for shear flow and heat convection. In the 1990s, Constantin and Doering introduced a practical framework for estimating physical quantities rigorously and directly from the equations of motion, which they called the ``background flow method'' \cite{DoeringConstantin94,
ConstantinDoering95,DoeringConstantin96}.

The background flow method is an extremely robust method for constructing bounds in fluid dynamics and builds up on techniques developed by Hopf to generalize Leray solutions of the Navier--Stokes equations to finite geometries with physical boundary conditions \cite{Hopf41}. In this method one manipulates the equations of motion relative to a steady trial background state. Decomposing the quantity of interest, e.g., the energy dissipation, into background profile, which satisfies the forcing conditions, and fluctuating component, the background part yields an upper bound if the fluctuation term satisfies a certain nonnegativity condition, which is often referred to as the spectral constraint. In a certain sense, finding the least upper bound using Constantin and Doering's method resembles a variational saddle point problem. The equations imposed on the fluctuations necessarily include the equations of motion. To simplify the derivation of the Euler--Lagrange equations corresponding to the variational problem for the fluctuation term, however, fluctuations are often chosen in a much larger class of functions. In other words, the spectral constraint in the background method is required to hold for an infinite-dimensional set of vector fields, that strictly contains the solutions of the equations of motion. In that case, enforcing the spectral constraint may possibly yield to an overestimation of the quantity of primal interest.

After its introduction, the background flow method started its triumph as {\em the} upper bound method with applications ranging from various problems in turbulent heat convection and boundary-force and body-force driven turbulence to idealized models in magnetohydrodynamics. The method was soon improved by Nicodemus {\em et al.} who introduced an additional balance parameter \cite{NicodemusGrossmannHolthaus97}, and Kerswell showed that this improved method is actually ``equivalent'' to the approaches of Busse and Howard \cite{Kerswell98}.

Apart from its practical performance, for many years, the background flow method was considered as a rigorous manifestation of Malkus's marginally stable boundary layer theory \cite{Malkus54}. The latter is based on the assumption that turbulent boundary driven flows organize themselves into marginally stable configurations. If the well-mixed core is bounded by thin laminar boundary layers, the thickness of these layers is determined by the condition of marginal stability. The association of the background flow method with Malkus's theory  relies on the surprising (?) observation that the spectral condition in the background flow method resembles a nonlinear stability condition on the background flow. A recent work of Nobili and Otto \cite{NobiliOtto14}, however, proves the failure of this association --- at least in the context of infinite Prandtl number Rayleigh--B\'enard convection: The authors compute the least upper bound on the Nusselt number (the quantity of interest in Rayleigh--B\'enard convection) within the framework of the background flow method. This bound however, exceeds the bound derived by Otto and the author in \cite{OttoSeis11} using completely different methods. In the context of Rayleigh--B\'enard convection, it thus seems that a physical interpretation of the background flow method is misleading. Whether the background flow method indeed gives physically relevant information (apart from scaling bounds) in different problems of fluid dynamics can only be speculated.

In this paper, we focus on the energy dissipation rate as an example of one specific physical quantity and present a new method for its rigorous\footnote{We caution the reader that our results are still only ``formally'' true in the sense that many of the performed mathematical operations apply only to sufficiently smooth solutions of the Navier--Stokes equations. Of course, the results can be made rigorous if the analysis is performed on suitable week solutions, e.g., Leray solutions, in the appropriate framework.} estimation directly from the equations of motion. (In fact, the method was already introduced in \cite{OttoSeis11} by Otto and the author, but its universality was not seen at that time.) To allow for a straight comparison with the background flow method, the method is tested on the problems considered by Constantin and Doering in \cite{DoeringConstantin94,
ConstantinDoering95,DoeringConstantin96,
DoeringConstantin98}. More precisely, we study the classical fluid dynamics problems {\em shear flow}, {\em channel flow}, {\em Rayleigh--B\'enard convection} and {\em porous medium convection}. 
All of these problems have in common that they can be considered as model problems for boundary-force or body-force driven flows or for thermal convection. In fact, we will recover the same results as Constantin and Doering in the above mentioned papers.

Our new approach presented in this paper is entirely different from the background flow method in that it is rather based on (local) conservation laws. More precisely, at the heart of our method conservation laws for the certain flux components are established. The conservation laws are local in the direction of the symmetry axis but global in all other spatial and time variables. Typically, the conserved quantity can be explicitly related to the energy dissipation rate. The former can be averaged over boundary layers along the rigid domain walls where the velocity field is small due to no-slip boundary conditions and it can thus be easily controlled by the viscous dissipation rate. Optimizing the width of the boundary layer then yields a bound on the energy dissipation rate. How the new method applies to fluids with different boundary conditions, e.g., no-stress, has to be seen.

The author believes that the strength of the new method relies on the fact that it uses the equations motions and some secondary derived physical laws directly instead of working with a rigid ``upper bound construction'' which in some cases is too restrictive to yield the optimal result, cf.\  \cite{OttoSeis11,NobiliOtto14}. The involved mathematical operations are elementary.

%
%
%
%
%
%The background flow method has proved to be a very powerful tool in the upper bound theory. Its applications are numerous and in many cases a suitable trial background flow could simply be chosen as a piecewise linear function. Yet, in many cases, the eigenvalue problem associated with the spectral constraint is in general not explicitly solvable. 
%
%

We apologize in advance that in this paper we will not apply any effort in explicitly computing nor optimizing the prefactors in our bounds. The aim of this paper is solely to advertise an alternative method to the most widely used background flow method to derive upper bounds on physical quantities in turbulent fluid flows.

The article is organized as follows: After fixing the notation in Section \ref{S2}, we will derive upper bounds on the energy dissipation for shear flows (Section \ref{S3}), channel flows (Section \ref{S4}), Rayleigh--B\'enard convection (Section \ref{S5}) and porous medium convection (Section \ref{S6}).

\section{Notation}\label{S2}
In what follows, we will be try to be as consistent as possible with regard to the notation even though different physical problems will be considered.

Our models are nondimensionalized.

Our system is a layer of fluid in the box $[0,L)^{d-1}\times [0,1]$ where $L$ is an arbitrary positive number that will not enter in our analysis. Throughout the article, we will refer to the first $d-1$ coordinates as the horizontal and the last coordinate as the vertical one. We write $x=(y,z)\in \R^{d-1}\times \R$ accordingly, and denote by $\{e_1,\dots, e_{d-1}, e_d\}$ the canonical basis of $\R^d=\R^{d-1}\times \R$.

We assume periodic boundary conditions in the horizontal directions for all variables involved. The horizontal boundaries are rigid and the imposed conditions will depend on the particular physical problem under consideration.

The fluid velocity is denoted by $u$, and we write $u=(v,w)\in \R^{d-1}\times \R$ to distinguish the horizontal velocity vector from the vertical component. The hydrodynamic pressure is $p$ and $T$ is the temperature field.

We consider the rate of energy dissipation
\[
\eps = \int_0^1 \la|\grad u|^2\ra\, dz,
\]
where $\la \cdot \ra$ denotes the horizontal and time average, that is,
\[
\la f\ra = \lim_{\tau \uparrow \infty}\frac{1}{\tau}\int_0^{\tau} \frac1{L^{d-1}} \int_{[0,L)^{d-1}} f(t,y)\, dydt.
\]
In general, the long-time average need not to exists, even if finite time averages are bounded, and we could be more careful at this point by choosing the $\limsup$ instead of the $\lim$. However, for the sake of a clearer statement and to simplify the subsequent analysis, we will be quite formal in most of our computations.

\section{Shear flow}\label{S3}

As the simplest example for a boundary-driven flow, we consider a fluid which is confined between two parallel plates that are moving at a constant speed relative to each other. The equations of motion are the Navier--Stokes equations in a box $[0,L)^{d-1}\times [0,1]$,
\begin{eqnarray}
\partial_t u + u\cdot \grad u  - \laplace u+\grad p&=&0,\label{1}\\
\div u&=& 0,\label{2}
\end{eqnarray}
and, as stated in the previous section, the velocity $u$ and the pressure $p$ are both periodic in the horizontal variables. At the horizontal plates, we assume no-slip boundary conditions for the velocity field. If the upper boundary plate is moving with constant speed $\Re$ in direction $e_1$ while the lower plate is at rest, we must have
\begin{equation}\label{3}
u = 0\mbox{ for }z=0\quad\mbox{and}\quad u=(\Re) e_1\mbox{ for }z=1.
\end{equation}
Notice that all quantities are nondimensionalized and $\Re$ is the Reynolds number.

The scaling of the energy dissipation rate $\eps$ as a function of the Reynolds number is of fundamental importance in many engineering applications, since for steady states, the energy dissipation rate measures the rate at which work must be done by an agent to keep the upper plate moving.

The interest in understanding the energy dissipation rate in boundary driven fluid flows as a function of the Reynolds number dates back to Stokes. It is well-known \cite{Serrin59,KellerRubenfeldMolyneux67} that solutions to the Stokes equation minimize the dissipation rate among all divergence-free vector fields with a fixed velocity at the boundary.
In many situations of physical interest, this solution is laminar and continues thus to exist as a solution of the Navier--Stokes equation. In our case, the laminar solution is the so-called Couette flow $u_C(z) = (\Re)z e_1$ which dissipates energy $\eps_C=(\Re)^2$. For small Reynolds numbers, $u_C$ is a stable solution of the Navier--Stokes equation in the sense that any sufficiently small perturbation will decrease in time. For large Reynolds numbers, however, perturbations of the laminar steady state are unstable and the flows can become chaotic or turbulent. The energy dissipation rate is a monotone function of the Reynolds number and acts as a measure of how turbulent the flow is. With regard to the presence of steady regular solutions also in the high Reynolds number regime---let us call these solutions ``ungeneric''---absolute lower bounds on the energy dissipation will be dictated by unturbulent flows: $\eps\ge\eps_C= (\Re)^2$. A rigorous scaling theory for the dissipation rate can hence only be an upper bound theory. An upper bound, however, sets limits on the possible turbulent structures of the flow and is thus an indispensable knowledge in the study of turbulence.

Developing a conventional statistical turbulence theory for high Reynolds numbers, Constantin and Doering predict the ``logarithmic friction law''
\[
\eps \sim \frac{(\Re)^3}{(\log \Re)^2}\quad\mbox{as }\Re\gg1,
\]
cf.\ \cite[Appendix A]{DoeringConstantin94}, which is in accordance with the experimental data derived in \cite{LathropFinebergSwinney92}. In the same paper, the authors derive a first upper bound which proves the conjectured rate up to the logarithmic factor, that is, they prove that $\eps\lesssim(\Re)^3$ using the background flow method. In the following, we reproduce Constantin and Doering's bound using our new method.

As a starting point of our analysis, we recall that the energy dissipation can equally be expressed, for instance, as the trace of the vertical derivative of $v_1= u\cdot e_1$ at the top plate, that is,
\begin{equation}\label{4}
\eps = \Re\la \left.\partial_z\right|_{z=1}v_1\ra.
\end{equation}
The identity follows from testing \eqref{1} with $u$, integrating by parts and using \eqref{2} and \eqref{3}. Moreover,
 multiplying \eqref{1} by $e_1$, taking the horizontal and time average, and using \eqref{2} and the periodicity, we see that
\[
\la w v_1 - \partial_z v_1\ra = \const\quad\mbox{for all }z\in[0,1],
\]
and thus, in view of \eqref{3} and \eqref{4}, we obtain the formula
\[
\eps = \Re\la \partial_z v_1-w v_1  \ra\quad\mbox{for all }z\in[0,1].
\]
In particular, averaging this identity over some boundary layer $[0,\ell]$, where $0\le \ell\le 1$ has to be determined later, we see that
\[
\eps = \frac{\Re}{\ell} \la \left.{v_1}\right|_{z=\ell}\ra - \frac{\Re}{\ell} \int_0^{\ell} \la w v_1\ra\, dz ,
\]
where we have used the fundamental theorem together with the no-slip boundary condition \eqref{3}. We now apply the Cauchy-Schwarz and Poincar\'e inequalities and deduce
\begin{eqnarray*}
\eps &\le&  \frac{\Re}{\ell} \sup_{z\in[0,\ell]} \la |u|^2\ra^{1/2}+\frac{\Re}{\ell} \int_0^{\ell} \la |u|^2\ra\, dz \\
&\le& \frac{\Re}{\ell^{1/2}}\left( \int_0^{\ell} \la |\partial_z u|^2\ra\, dz\right)^{1/2}+(\Re) \ell \int_0^{\ell} \la |\partial_z u|^2\ra\, dz  \\
&\le&  \frac{\Re}{\ell^{1/2}} \eps^{1/2}+(\Re) \ell \eps .
\end{eqnarray*}
Optimizing the last expression in $\ell$ yields $\ell\sim \eps^{-1/3}\ll1$, and thus $\eps\lesssim (\Re) \eps^{2/3}$, which entails
\[
\eps\lesssim (\Re)^3.
\]
This estimate agrees with the bound derived in \cite{DoeringConstantin94}.
% Some elements of this approach were already used in \cite{GebhardtGrossmannHolthausLoehden95}

\section{Channel flow}\label{S4}

As the simplest example for a body-force driven flow, we consider a fluid in a rectangular domain, which is driven by a pressure gradient in a direction of one of the horizontal boundary plates. The problem is modelled by the forced Navier--Stokes equation
\begin{eqnarray}\label{11}
\partial_t u + u\cdot \grad u - \laplace u + \grad p &=& (\Gr) e_1,\\
\div u&=& 0,\label{11a}
\end{eqnarray}
in $[0,L)^{d-1}\times [0,1]$, supplemented with no-slip boundary conditions
\begin{equation}\label{11b}
u=0\quad\mbox{for }z\in\{0,1\}.
\end{equation}
Here, $\Gr$ denotes the nondimensional Grashof number. We remark that when the force is specified a priori as in the present situation, the Reynold number is an emergent quantity.

The channel flow problem has the laminar solution $u_P(z) = \frac12 (\Gr) (z^2-z)e_1$, the so-called Poiseuille flow, for which the energy dissipation rate is $\eps_P = (\Gr)^2/12$. The Poiseuille flow plays a similar role in the channel flow problem as the Couette flow in the shear flow problem: $u_P$ solves the corresponding Stokes equation and is an unstable solution of the Navier--Stokes equation in the high Grashof number regime. When bounds are expressed in terms of the Grashof number instead of the Reynolds number, the Stokes limit represents an upper bound on the energy dissipation rate: $\eps\le \eps_P = (Gr)^2/12$.
 
The scaling of the energy dissipation rate in the high Grashof number regime is expected to obey the modified logarithmic friction law
\[
\eps \sim \frac{(\Gr)^{3/2}}{(\log \Gr)^2}
\]
--- similar to the shear flow. Due to the presence of ungeneric solutions of the Navier--Stokes equation, we can only expect to derive a one-sided version of this scaling law. A first rigorous lower bound on the energy dissipation rate was established by Constantin and Doering \cite{ConstantinDoering95}, $\eps\gtrsim (\Gr)^{3/2}$, which is optimal up to the logarithm. In the following, we will recover this bound following our new method. For further improvements we refer to \cite{PetrovLuDoering05} and references therein.

We see from testing \eqref{11} with $u$, integrating by parts and using \eqref{11a} and \eqref{11b} that the energy dissipation rate is proportional to the average flow velocity
\begin{equation}\label{13}
\eps = \Gr \int_0^1 \la v_1\ra\, dz.
\end{equation}
Moreover, multiplying \eqref{11} by $e_1$, taking the horizontal and time average, and using \eqref{11a} and periodicity, we obtain
\[
\partial_z \la w v_1 -\partial_z v_1\ra = \Gr\quad\mbox{for all }z\in[0,1].
\]
From the no-slip boundary conditions \eqref{11b} we thus infer that
\begin{equation}\label{12}
\la wv_1 - \partial_z v_1\ra = (\Gr) z - \la \left.\partial_z\right|_{z=0} v_1\ra \quad\mbox{for all }z\in[0,1].
\end{equation}
On the one hand, averaging in $z$ over the strip $[1-\ell,1]$ of width $\ell\ll1$ yields
\[
\Gr \sim \frac1{\ell} \int_{1-\ell}^1 \la v_1w -\partial_z v_1\ra\, dz + \la \left.\partial_z\right|_{z=0} v_1\ra .
\]
On the other hand, averaging \eqref{12} over $[0,\ell]$ yields
\[
\la \left.\partial_z\right|_{z=0} v_1\ra \sim \frac1{\ell} \int_0^{\ell}\la\partial_z v_1- v_1w  \ra\, dz  + \ell \Gr.
\]
Since $\ell\ll1$, a combination of the previous two estimates gives
\[
\Gr \lesssim \frac1{\ell} \int_0^{\ell} \la  \partial_z v_1-  v_1w\ra\, dz +\frac1{\ell} \int_{1-\ell}^1 \la v_1w -\partial_z v_1\ra\, dz.
\]
We can apply the same arguments as in the shear flow case considered in the previous section to deduce
\[
\Gr \lesssim \frac{\eps^{1/2}}{\ell^{1/2}} + \ell\eps.
\]
The optimal $\ell$ is of size $\ell \sim \eps^{-1/3}$, which implies that
\[
(\Gr)^{3/2} \lesssim \eps.
\]
In view of \eqref{13}, this bound is equivalent to
\[
(\Gr)^{1/2} \lesssim \int_0^1 \la v_1\ra\, dz,
\]
which agrees with the bound derived in \cite{ConstantinDoering95} and appears to be sharp to within logarithms.

\section{Rayleigh--B\'enard convection}\label{S5}

Rayleigh--B\'enard convection is the transport of heat by thermal convection in a fluid layer that is heated from below and cooled from above. The problem is modelled by the equations of the Boussinesq approximation
\begin{eqnarray}
\partial_t T  + u\cdot \grad T - \laplace T &=&  0,\label{5a}\\
\div u&=& 0\label{5b},\\
\frac1{\Pr} \left(\partial_t u+ u\cdot \grad u\right) -\laplace u + \grad p &=& \Ra T e_d.\label{5c}
\end{eqnarray}
The system is nondimensionalized and admits two controlling parameters, the Rayleigh number $\Ra$ and the Prandtl number $\Pr$. The equations are complemented by the boundary conditions
\begin{equation}
\label{5d}
T = 1\quad\mbox{ on }z=0,\quad\mbox{and}\quad T = 0\quad\mbox{ on }z=1,
\end{equation}
representing heating at the bottom and cooling at the top, and no-slip boundary conditions for the velocity field,
\begin{equation}
\label{5e}
u=0\quad\mbox{ on }z\in\{0,1\}.
\end{equation}
The quantity of interest in this model is the so-called Nusselt number, a measure for the average upward heat flux. It is defined by
\[
\Nu = \int_0^1\la \left(uT-\grad T\right)\cdot e_d\ra\, dz  = \int_0^1 \la wT - \partial_z T\ra\ dz.
\]
The scaling of the Nusselt number in terms of $\Ra$ and $\Pr$ is a problem of enormous experimental, numerical and theoretical research for over fifty years. For a recent review, we refer to \cite{AhlersGrossmannLohse09} and references therein. For all values of $\Ra $ and $\Pr$, a laminar solution is given by $u_c=0$ and $T_c=1-z$, which corresponds to pure conduction. The corresponding Nusselt number is $\Nu_c =1$, and the laminar solution is unstable for large Rayleigh numbers.

In the high Rayleigh number regime, $\Ra\gg1$, the scaling of the Nusselt number is proportional to the scaling of the energy dissipation rate. Indeed, testing \eqref{5c} with $u$, using the imcompressibility assumption \eqref{5b} and invoking the boundary conditions \eqref{5d} and \eqref{5e} for $T$ and $u$ yields that
\begin{equation}\label{5ca}
\eps = \Ra(\Nu-1)\sim \Ra \Nu.
\end{equation}
The bound on the Nusselt number in the ultimate turbulent regime is expected to be
\[
\Nu  \sim \Ra^{1/2},
\]
if $\Ra\gg1$, uniformly in $\Pr$.\footnote{In fact, this scaling is in conflict with the rigorous bound derived by Whitehead and Doering for 2D Rayleigh--B\'enard convection with free-stress boundary conditions, in which case $\Nu \lesssim \Ra^{5/12}$ could be proved \cite{WhiteheadDoering11}. The proof makes heavily use of the 2D structure (no vortex stretching) and the free-slip boundary conditions (homogeneous vorticity boundary conditions). Whether such a bound can be extended to our problem at hand is not clear to the author. Recent numerical simulations at least indicate that the $5/12$ scaling should be expected for any solution of \eqref{5a}--\eqref{5c} with finite energy dissipation rate \cite{HassanzadehChiniDoering14}.}
In the following, we derive an upper bound on this scaling with the help of different representations of the Nusselt number, similar to the approach in the previous two sections. Averaging the heat equation \eqref{5a} and using periodicity and \eqref{5b}, it follows that the heat flux is constant on every horizontal slice, that is
\[
\Nu = \la Tw - \partial_z T\ra\quad\mbox{for all }z\in[0,1].
\]
Now, averaging this identity over a boundary layer of thickness $\ell\in[0,1]$, and using the maximum principle on the temperature $\max |T|\le 1$, which is enforced by the boundary conditions (if not initially, then exponentially fast in time), we obtain that
\[
\Nu\le \frac1{\ell} \int_0^{\ell} \la wT\ra\, dz +\frac1{\ell} \le \frac1{\ell}\int_0^{\ell} \la |w|\ra\, dz + \frac1{\ell}.
\]
We use Poincar\'e's and H\"older's inequalities and \eqref{5e} to bound the integral over the vertical velocity component by the energy dissipation rate $\eps$, that is
\[
\int_0^{\ell} \la |w|\ra\, dz\le \ell \int_0^{\ell} \la  |\partial_z w|\ra\, dz\le \ell^{3/2} \left(\int_0^{\ell} \la (\partial_z w)^2\ra\, dz \right)^{1/2} \le \ell^{3/2} \left( \Ra Nu\right)^{1/2}
\]
by \eqref{5ca}, so that
\[
\Nu \le \ell^{1/2} \left(\Ra \Nu\right)^{1/2} + \frac{1}{\ell}.
\]
Optimizing in $\ell$ yields that $\ell \sim \left(\Ra\Nu\right)^{-1/3}$, which entails that
\[
\Nu\lesssim \Ra^{1/2}.
\]
This is precisely the same scaling law derived by Constantin and Doering in \cite{DoeringConstantin96}.

Applying the same method combined with sophisticated maximal regularity arguments, Choffrut, Nobili and Otto \cite{ChuffrutNobiliOtto14} recently obtained new bounds on $\Nu$ which improve this bound in certain $\Ra$-$\Pr$ regimes. The results in particular apply to the large Prandtl number regime. The Nusselt number bound can be interpreted as a bound on the average temperature gradient, cf.\ \eqref{14} below. Developing techniques similar to those presented in this paper, the author derived bounds on higher order derivatives of the temperature field in infinite Prandtl number convection and estimated deviations of the average vertical temperature profile from linearity \cite{Seis13}.

%A qualitatively new stage in the upper bound theory was reached later by the same authors \cite{ConstantinDoering99} who incorporated ideas from Calderon--Zygmund maximal regularity theory in their mathematically fully rigorous analysis of the Nusselt number scaling in Rayleigh--B\'enard convection, which is proportional to the the energy dissipation rate.
%

\section{Porous medium convection}\label{S6}

We finally consider thermal convection in a porous medium. In this case, Darcy's law approximates the Navier--Stokes equations, and the Rayleigh--B\'enard system \eqref{5a}--\eqref{5c} reduces to
\begin{eqnarray}
\partial_t T  + u\cdot \grad T-\laplace T&=&0,\label{6}\\
\div u&=&0,\label{7}\\
u+\grad p &=& (\Ra) Te_d.\label{8}
\end{eqnarray}
As before, the non dimensional number $\Ra$ is the Rayleigh number. The boundary condition satisfied by the fluid velocity are
\begin{equation}\label{9}
w=0\quad \mbox{on }z\in \{0,1\},
\end{equation}
and we suppose that the container is cooled from above and heated from below, modeled by
\begin{equation}\label{10}
T=1\quad\mbox{on }z=0,\quad\mbox{and}\quad T=1\quad\mbox{on }z=1.
\end{equation}
Again, we assume periodic boundary conditions in all horizontal directions for all quantities involved.

As in the case of classical Rayleigh--B\'enard convection, the quantity of interest here is the Nusselt number
\[
\Nu = \int_0^1\la wT-\partial_z T \ra\, dz.
\]
Before estimating  the energy dissipation rate in this example, we start with the study of the Nusselt number. 
Experiments and numerics suggest that
\[
\Nu\sim \Ra\quad\mbox{as }\Ra\gg1,
\]
cf.\ \cite{HewittNeufeldLister12}. Because of the existence of ungeneric laminar solutions, we can only expect to prove the upper bound $\Nu\lesssim \Ra$.
We first need to establish some alternative identifies for $\Nu$. We first recall that 
averaging \eqref{6}, we obtain $\la wT - \partial_z T\ra=\const$, and thus
\[
\Nu = \la wT - \partial_z T\ra\quad\mbox{for all }z\in[0,1].
\]
In particular, $\Nu = -\la \left.\partial_z\right|_{z=0} T\ra$. Testing now \eqref{6} with $T$, integrating by parts yields and using \eqref{7}, \eqref{9} and \eqref{10}, we see that
\begin{equation}
\label{14}
\Nu = \int_0^1\la |\grad T|^2\ra\, dz.
\end{equation}
On the other hand, by the definition of $\Nu$ and \eqref{10}, testing \eqref{8} with $u$ and using \eqref{7} and \eqref{9}, we obtain
\begin{equation}\label{14a}
Nu = \frac1{\Ra} \int_0^1\la |u|^2\ra\, dz +1.
\end{equation}

We are now able to estimate the Nusselt number. Letting $\ell\in(0,1)$ be an arbitrary number, we estimate
\[
\Nu\le \frac1{\ell}\int_0^{\ell} \la wT\ra\, dz + \frac1{\ell}
\]
as in the previous section. Because $|T|\le 1$ by the maximum principle for the temperature, we obtain via Jensen's inequality,
\[
\Nu \le \left(\frac1{\ell}\int_0^{\ell} \la w^2\ra\, dz\right)^{1/2} + \frac1{\ell}.
\]
By the Nusselt number representation \eqref{14a}, this yields
\[
Nu \lesssim \frac{ (\Ra \Nu)^{1/2}}{\ell^{1/2}}  + \frac1{\ell}.
\]
This bound is optimized by $\ell\sim 1$, so that
\[
\Nu\lesssim \Ra
\]
because $\Ra\gg1$.
%\footnote{We like to point out that the use of the maximum principle in this argument is not essential. An alternative argument is by estimating $\Nu$ in the first step with the help of the Cauchy--Schwarz inequality and controlling the $L^2$ norm of $T$ by $\Nu$ via Poincar\'e's inequality.}

In a final step, we like to relate the current bound on the Nusselt number to a bound on the viscous dissipation rate. In fact, we will show that $\eps\lesssim (\Ra)^2 \Nu$, so that the above statement turns into
\[
\eps\lesssim (\Ra)^3.
\]
Indeed, we start noting that Darcy's law \eqref{8} together with the boundary conditions \eqref{9} and \eqref{10} provides us with Neumann boundary conditions for the pressure: $\left.\partial_z\right|_{z=0} p =\Ra$ and $\left.\partial_z\right|_{z=1} p =0$. In particular, differentiating the horizontal velocity components with respect to $z$, $\partial_z v = -\grad_y\partial_z p$, multiplying by $v$, averaging and integrating by parts yields
\[
\la v\cdot \partial_z v\ra = -\la v\cdot \grad_y\partial_z p\ra = \la (\grad_y\cdot v)\partial_z p\ra.
\]
In particular, the above values for $\partial_z p$ and the horizontal periodicity imply that
\begin{equation}\label{14b}
\left.\la v\cdot \partial_z v\ra \right|_{z=0,1}  = 0.
\end{equation}
It thus follows via integration by parts
\begin{equation}\label{14c}
\int_0^1 \la |\grad v|^2\ra\, dz = \left.\la v\cdot \partial_z v\ra \right|^{z=1}_{z=0} - \int_0^1 \la v\cdot \laplace v\ra\, dz  \stackrel{\eqref{14b}}{=} - \int_0^1 \la v\cdot \laplace v\ra\, dz.
\end{equation}
Now notice that $-\laplace v = (\Ra) \grad_y\partial_z T$. Indeed, taking the divergence of \eqref{8} yields $\laplace p = (\Ra) \partial_z T$, and thus $-\laplace v = \grad_y \laplace p = (\Ra) \grad_y \partial_z T$. Therefore, \eqref{14c} becomes
\[
\int_0^1 \la |\grad v|^2\ra \, dz = \Ra \int_0^1 \la v\cdot \grad_y \partial_z T\ra\,dz = -\Ra \int_0^1 \la ( \grad_y\cdot v) \partial_z T\ra\, dz.
\]
Since $|\grad_y\cdot v|\lesssim |\grad_y v|\le |\grad v|$, we can use the Cauchy--Schwarz inequality to obtain
\[
\int_0^1 \la |\grad v|^2\ra\, dz \lesssim \Ra\left(\int_0^1 \la|\grad v|^2\ra\, dz\int_0^1 \la|\grad T|^2\ra\, dz\right)^{1/2},
\]
and thus, via \eqref{14},
\[
\int_0^1 \la |\grad v|^2\ra\, dz \lesssim (\Ra)^2 \Nu.
\]
The estimate of the vertical velocity component is easier because of \eqref{9}. Indeed, testing $\laplace w=(\Ra) \laplace T - \partial_z \laplace p  = (\Ra) \laplace_y T$ with $w$, integrating by parts and using Young's inequality yields
\[
\int_0^1 \la |\grad w|^2\ra\,dz\lesssim (\Ra)^2\Nu.
\]
Combining the last two estimates finally yields $\eps\lesssim (\Ra)^2\Nu$ as desired.

\section*{Acknowledgment}
Part of this work was performed when the author was visiting the MPI MIS in Leipzig. The author is grateful for the hospitality. He further acknowledges stimulating discussions with Camilla Nobili, Charlie Doering, Jared Whitehead and Ralf Wittenberg. 

\bibliography{rbc_lit}
\bibliographystyle{abbrv}
\end{document}